\documentclass[english,aps,preprint]{revtex4}
\usepackage[T1]{fontenc}
\usepackage{array}
\usepackage{amsmath}
\usepackage{amssymb}
\usepackage{graphicx}

\makeatletter

\providecommand{\tabularnewline}{\\}
\makeatother
\usepackage{babel}

\begin{document}

\title{Comparing the Topological and Electrical Structure of the North American
Electric Power Infrastructure}

\author{Eduardo Cotilla-Sanchez }

\email{eduardo.cotilla-sanchez@uvm.edu}

\author{Paul D. H. Hines }

\email{paul.hines@uvm.edu}

\affiliation{School of Engineering, University of Vermont, Burlington, VT 05405,
USA}

\author{Clayton Barrows }

\email{cpb155@psu.edu}

\author{Seth Blumsack}

\email{sethb@psu.edu}

\affiliation{John and Willie Leone Family Department of Energy and Mineral Engineering,
The Pennsylvania State University, University Park, PA 16802}

\begin{abstract}
The topological (graph) structure of complex networks often provides
valuable information about the performance and vulnerability of the
network. However, there are multiple ways to represent a given network
as a graph. Electric power transmission and distribution networks
have a topological structure that is straightforward to represent
and analyze as a graph. However, simple graph models neglect the comprehensive
connections between components that result from Ohm's and Kirchhoff's
laws. This paper describes the structure of the three North American
electric power interconnections, from the perspective of both topological
and electrical connectivity. We compare the simple topology of these
networks with that of random \cite{Erdos:1959}, preferential-attachment
\cite{Barabasi:1999} and small-world \cite{Watts:1998} networks
of equivalent sizes and find that power grids differ substantially
from these abstract models in degree distribution, clustering, diameter
and assortativity, and thus conclude that these topological forms
may be misleading as models of power systems. To study the electrical
connectivity of power systems, we propose a new method for representing
electrical structure using electrical distances rather than geographic
connections. Comparisons of these two representations of the North
American power networks reveal notable differences between the electrical
and topological structure of electric power networks.\end{abstract}
\maketitle
\section{Introduction}

Recent research in complex networks \cite{Boccaletti:2006}
has elucidated strong links between network structure and performance.
Scale-free networks, which are characterized by strongly heterogeneous
(power-law) node connectivity (degree), are uniquely robust to random
failures but vulnerable to directed attacks \cite{Barabasi:1999,Albert:2000}.
Graphs with exponential degree distributions, such as the random graph
\cite{Erdos:1959} and small-world networks \cite{Watts:1998} are
more equally vulnerable to random failures and directed attacks. Scale-free
networks also tend to lose synchronization when attacked at hub nodes
\cite{Wang:2002}, which is not the case for random graphs. And, for
controllable networks, scale-free networks can be synchronized by
controlling a small number of highly connected nodes \cite{Li:2004},
or even a single node \cite{Chen:2007}. On the other hand, ref. \cite{Atay:2006}
shows that many classes of networks fail to synchronize, arguing that
degree distribution alone is not sufficient to characterize the performance
of a network. References \cite{Wu:1995,Wu:2005} describe how network
structure influences consensus (a form of synchronization), and how
different ways of representing a system a graph may affect the conclusions
that one draws about performance. Others~\cite{De-Lellis:2008,De-Lellis:2010a}
show that it is possible to maintain synchronization in evolving network
topologies. 

Given that network structure can dramatically influence performance,
and given the size, complexity and importance of electric power systems,
it is not surprising that power grids are the subject of substantial
study from a complex networks perspective \cite{Dorfler:2010}. The
fact that data from many countries show a power-law in power system
blackout sizes \cite{Dobson:2007}, leads naturally to the conjecture
that this might result from complexity in network structure. Watts
and Strogatz \cite{Watts:1998} measure characteristic path length
and clustering in power grids, and find similarities to the small-world
network model. A number of studies measure the degree distribution
of various power grids with some reporting exponential \cite{Amaral:2000,Albert:2004,Crucitti:2004}
and others reporting power-law/scale-free degree distributions \cite{Barabasi:1999,Chassin:2005}.
That studies of different power grids in different countries or regions
yield different topological structures is not necessarily surprising.
More surprising is that different analyses of identical grids (the
transmission system in the Western U.S.) have yielded different structural
results \cite{Albert:2004,Chassin:2005}. In particular, this discrepancy
stems from the model used in \cite{Chassin:2005} which circumvents
the exponential cut-off reported in \cite{Amaral:2000} by adding
synthetic distribution nodes. This paper clarifies this uncertainty
by showing that, at least for the IEEE 300 bus network and the US
Eastern Interconnect, an exponential degree distribution is a better
fit to the data than a power-law distribution. In addition to reporting
an exponential degree distribution, ref. \cite{Albert:2004} also
reports a power-law in the topological betweenness of nodes in power
grids, which is proposed as a potential explanation for the power-law
blackout frequencies. Reference \cite{Holmgren:2006} compares the
Western US power grid and the Nordic grid using the topological attack/failure
model in \cite{Albert:2000}, reports that the Nordic grid has a more
heavy-tailed degree distribution structure than the Western grid,
and provides some recommendations for increasing the robustness of
power grids. Reference\ \cite{Sole:2008} studies the European power
grid and finds a positive correlation between topological robustness
measures and real non-topological reliability measures. Reference~\cite{Wang:2010}
uses results from a topological analysis to design a method for generating
synthetic power grids.

While they provide some insight into network structure, these topological
studies largely neglect Ohm\textquoteright{}s and Kirchhoff\textquoteright{}s
laws, which govern flows in electric circuits. In some cases pure
topological models can lead to provocative, yet erroneous, conclusions
about network vulnerability \cite{Wang:2009,Hines:2010b,Wang:2010a}.
To study electrical structure, refs.~\cite{Hines:2008,Hines:2010a}
describe a measure of Electrical Centrality for power grids. Bompard
et al.~\cite{Bompard:2009,Arianos:2009} combine topological models
with linearized power system models and propose new measures that
can be useful in identifying critical components. Also using a power-flow
model, ref.~\cite{Blumsack:2007a} identifies relationships between
Wheatstone structures within power grids, reliability and efficiency.

\subsection{Goals and outline of this paper}

This paper aims to fill a number of gaps in the existing research
on the structure of electric power systems by identifying similarities
and differences among the topological and electrical structure of
power grids and synthetic networks. The data used for this study include
several IEEE test cases, a 41,228 bus model of the US Eastern Interconnect
(EI), a 11,432 bus model of the US Western Interconnect (WI), and
a 4,513 bus model of the US Texas Interconnect (TI). These models
are substantially more detailed, and accurate, than models that have
been used in past structural studies of the North American power grid,
including previous work by the authors \cite{Hines:2010a}.

Section \ref{sec:topology} provides a topological analysis of the
two networks, showing how power grids differ substantially from random
\cite{Erdos:1959}, small world \cite{Watts:1998}, and preferential
attachment \cite{Barabasi:1999} graphs. Section \ref{sec:electrical}
proposes a new method for studying power grids as complex networks,
based primarily on electrical, rather than topological, structure.
Section \ref{sec:conclusions} discusses some implications of these
results for future studies of power grid performance.

\section{The topological structure of power grids\label{sec:topology}}

The existing literature shows some disagreement over the topological
structure of power networks. Some report a power-law degree distribution
\cite{Chassin:2005,Crucitti:2004} whereas others argue that an exponential
fit is superior \cite{Amaral:2000,Albert:2004}. Several report that
power grids have a small-world structure \cite{Watts:1998,Holmgren:2006,Ming:2006,Rosato:2007,Fu:2010},
from limited quantities of data. Some report that while the degree
distribution is exponential (such as is the case with random graphs),
power systems share properties with scale-free networks \cite{Albert:2004,Rosas-Casals:2007,Hines:2008}.
This section aims to clarify this and other uncertainties regarding
the structure of power grids using a larger, more accurate model of
the North American power grid than has been used in past studies.
To determine the extent to which power networks are similar to or
differ from common abstract network models, we compare the topology
of the three US interconnections (West, East, and Texas) to similarly-sized
small-world, scale-free and random graphs. This comparison builds
on that in \cite{Wang:2010} by including a more comprehensive statistical
description of not only power grids topology but also their counterpart
canonical graphs and by utilizing a larger, more accurate representation
of the North American power grid. Also, whereas ref. \cite{Wang:2010}
focused on the generation of random test networks, our goal is to
precisely describe the differences between topological and electrical
structure. 

To perform these comparisons, we represent each test network as an
undirected, unweighted graph with $n$ vertices/nodes and $m$ links/edges.
For the power grid models all buses, whether generator, load, or pass-through,
are treated equally. In this representation, links can represent two
or more parallel transformers or transmission lines, which means that
$m$ may be slightly smaller than the number of branches in the power
system model. The set of all vertices and links in each graph, $G$,
is $\{N,M\}$. The adjacency matrix for graph $G_{X}$ is $\mathbf{A}_{X}$,
with elements $a_{ij}$. 

The following sub-sections describe the power grid data, the synthetic
network models, and the metrics that are used for comparison purposes.

\subsection{Power grid data}

The data used in this study include several standard test power systems
(available from \cite{PSTCA:2007}), and a large model of the North
American power grid composed by the US Eastern Interconnect (EI),
US Western Interconnect (WI) and US Texas (ERCOT) Interconnect (TI).
The Eastern Interconnect (EI) data come from a North American Electric
Reliability Corporation (NERC) planning model for 2012. The Western
and Texas data come from the FERC form 715 filings from 2005.%
\footnote{The US power grid data used in this paper were obtained through the
U.S. Critical Energy Infrastructure Information (CEII) request process
(Federal Energy Regulatory Commission, Online: http://www.ferc.gov/legal/ceii-foia/ceii.asp).%
}

The IEEE 300 bus test system has 411 branches, and average degree
($<k>=2m/n$) of 2.73. Two of the branches are parallel links, so
the graph size is: $|G_{300}|=\{300,409\}$. After removing isolate
buses and parallel branches from the EI data we obtain a graph ($G_{EI}$)
with 41,228 vertices, 52,075 links and an average degree $<k>=2.53$.
After applying the same process to $G_{WI}$ we obtain a graph with
11,432 vertices, 13,734 links and an average degree $<k>=2.40$. For
the Texas Interconnection $|G_{TI}|=\{4513,5532\}$ and $<k>=2.45$.

\subsection{Synthetic networks}

In order to compare power grids to common graph structures, we synthesized
similarly sized graphs using the small-world \cite{Watts:1998}, preferential
attachment (scale-free) \cite{Barabasi:1999}, and random \cite{Erdos:1959}
graph models, each of which is described below.

The early work on small-world networks \cite{Watts:1998} argues that
power networks have the properties of a small-world system. Others
also report this to be true \cite{Holmgren:2006,Ming:2006,Rosato:2007,Fu:2010}.
To test whether this is indeed the case we generate a regular lattice
with $n$ nodes and approximately $m$ links. Nodes and links are
initially laid out to form a ring lattice, by forming links to neighboring
nodes, such that each node $a$ is connected $a\leftrightarrow a-1$
($\forall a>1$). A second link is then created from node $a$ to
node $a-2$ (for $a>2$) with probability $\Pr(a\leftrightarrow a-2)=m/n-1$,
giving approximately $m$ links in total. After generating the regular
lattice in this manner, random re-wiring proceeds for each node with
probability $p$ as in \cite{Watts:1998}. We adjust $p$ to obtain
a network with the same diameter as the corresponding power grid. 

Similarly, several \cite{Chassin:2005,Crucitti:2004} argue that power
grids have a scale-free structure, as evidenced by a power-law in
the degree distribution. As with the small-world network we synthesize
scale-free networks to match the size of each power network. To create
a preferential attachment/scale-free (PA) graph with roughly $n$
nodes and $m$ links we modify the algorithm described in \cite{Barabasi:1999}
to allow for a fractional average degree $<k>$. For each new node
$a$ we initially add one link between $a$ and an existing node $b$
using the standard roulette wheel method. Specifically, node $b$
is selected randomly from the probability distribution $\Pr(a\leftrightarrow b)=k_{b}/\sum_{c=1}^{n}k_{c}$,
where $k_{c}$ is the degree at node $c$. After adding this initial
link a second is added with probability $m/n-1$. Thus the addition
of each new node results in an average of $1+(m/n-1)=m/n$ new links,
producing a preferential attachment graph with $n$ nodes and roughly
$m$ links. 

Finally, we synthesize random graphs by randomly linking node pairs
until there are exactly $m$ links in the system.

\subsection{Measures of graph structure}

If one of the above abstract models is a good model for power networks,
then statistical measures for the synthetic graphs should be similar
to those that come from the power networks. There are many useful
statistical measures for graphs. Among the most useful are degree
distribution \cite{Barabasi:1999}, characteristic path length \cite{Watts:1998},
graph diameter \cite{Albert:2002}, clustering coefficient \cite{Watts:1998}
and degree assortativity \cite{Barabasi:1999}. These measures provide
a useful set of statistics for comparing power grids with other graph
structures.

The probability mass function (pmf) for node connectivity, or degree
distribution, describes the diversity of connectivity in a graph.
While random graphs have an exponential (single scale) degree distribution,
many real networks have a power-law degree distribution. These scale-free
networks tend to have highly connected hubs, which can make the network
vulnerable to directed attack \cite{Albert:2000}. The degree of node
$i$ in a graph with adjacency matrix $A$ is:
\begin{equation}
k_{i}=\sum_{j=1}^{n}a_{ij}\label{eq:nodedegree}
\end{equation}
and the degree distribution is $\Pr(k=x)=n_{k}/n$, where $n_{k}$
is the number of nodes with degree $k$. The complementary cumulative
distribution function (ccdf) of degree in a scale-free network will
follow a power-law function: 
\begin{equation}
\Pr(k\geq x)=\left(\frac{x}{x_{\min}}\right)^{-\alpha+1}\label{eq:nodedegreeccdf}
\end{equation}
where $x_{\min}$ is a minimum value for the power-law tail of the
degree distribution. If the degree distribution is exponential, a
minimum value Weibull distribution provides a better fit to the data:
\begin{equation}
\Pr(k\geq x)=e^{-\left(\frac{x-x_{\min}}{\lambda}\right)^{\beta}.}\label{eq:weibull}
\end{equation}

While the size of a network can be measured by the number of nodes,
$n$ does not give much information about distances within the network,
which are often an indicator of network performance. Two measures
of network distance are commonly employed: diameter ($d_{\max}$)
and characteristic path length ($L$). If $\mathbf{D}$ is a distance
matrix, in which the off-diagonal elements $d_{ij}$ give the minimum
number of links that one would need to traverse to get from node $i$
to node $j$, then the diameter of the network is: 
\begin{equation}
d_{\max}=\max_{ij}d_{ij}.\label{eq:dmax}
\end{equation}
The characteristic path length ($L$) is the average of all $d_{ij}$:
\begin{equation}
L=\frac{1}{n(n-1)}\sum_{\substack{\forall i,j\\
i\neq j
}
}d_{ij}\label{eq:avshortestpath}
\end{equation}
Additionally, Section \ref{sec:electrical} compares several networks
using average nodal distance ($<d_{i}>$) from each node $i$ to other
nodes in the network:

\begin{equation}
<d_{i}>=\sum_{j=1}^{n}\frac{d_{ij}}{n-1}\label{eq:avtopdistance}
\end{equation}
The set of distance metrics described above are referred to as \emph{topological
distances }within the next sections of this paper.

The manner in which a type of network increases in diameter ($L$
or $d_{\max}$) is a useful indicator of structure. In small-world
networks and random graphs, $L$ increases roughly with $\ln n$ \cite{Watts:1998},
which gives these networks their characteristically small distances
between vertices. In a circular lattice $L$ increases linearly with
$n$, and in a regular 2-dimensional grid $L$ will increase with
$\sqrt{n}$. Small-world networks differ from random graphs in that
nodes in small-world graphs are highly clustered, as they are in regular
lattice or grid structures \cite{Watts:1998}. Therefore if power
networks were small-world in structure, we would expect to see high
clustering and small distances. Section \ref{sub:TopologicalResults}
describes an asymptotic analysis of $L$ for several power grids,
which indicates that power grids fall somewhere between small-world
and regular network structures. In this paper clustering is measured
with the clustering coefficient described in \cite{Watts:1998}:
\begin{equation}
C=\frac{1}{n}\sum_{i=1}^{n}c_{i}\label{eq:clusteringcoef}
\end{equation}
where the clustering of node $i$ ($c_{i}$) is

\begin{equation}
c_{i}=\frac{e_{i}}{(k_{i}(k_{i}-1))/2}\label{eq:indivclusteringcoef}
\end{equation}
and $e_{i}$ is the number of links within the cluster of nodes including
node $i$ and its immediate neighbors $N_{i}$:
\begin{equation}
e_{i}=\sum_{\forall j,k\in\{N_{i}\cup i\}}a_{jk}/2\label{eq:clusteringtriangles}
\end{equation}

Finally we compare the degree assortativity in the test networks.
Degree assortativity ($r$) in a network is defined in \cite{Newman:2002}
as the extent to which nodes connect to nodes with similar degree.
Formally, assortativity is the correlation in degree for the nodes
on opposite ends of each link \cite{Newman:2003a}:

\begin{equation}
r=\frac{m^{-1}\sum_{i=1}^{m}j_{i}k_{i}-\left[m^{-1}\sum_{i=1}^{m}\frac{1}{2}(j_{i}+k_{i})\right]^{2}}{m^{-1}\sum_{i=1}^{m}\frac{1}{2}(j_{i}^{2}+k_{i}^{2})-\left[m^{-1}\sum_{i=1}^{m}\frac{1}{2}(j_{i}+k_{i})\right]^{2}}\label{eq:assortativity}
\end{equation}
where $m$ is the number of links in the network and $j_{i}$/$k_{i}$
are the degrees of the endpoints of link $i$.

\subsection{Topological results\label{sub:TopologicalResults}}

Our analysis of the IEEE 300 bus and the North American power grid
models clearly indicates that power networks are neither small world,
nor scale-free in structure. Two hypothesis tests regarding degree
distribution provide evidence for this conclusion. Hypothesis~1 is
that the power networks have the same degree distribution as the same-sized
synthetic network ($\Pr(k)\sim\Pr(k:\mbox{EI})$). Hypothesis~2 is
that various networks have a power-law degree distribution ($\Pr(k)\sim k^{-\alpha}$).
We use a Kolmogorov-Smirnov t-test to evaluate each hypothesis. For
hypothesis~2, we determine the power-law distribution fit parameters
($\alpha$ and $x_{\max}$) using the method described in \cite{Clauset:2009}.
Table \ref{tab:Network-stats} shows results from these tests, as
well as other measures of network structure. 

\begin{table*}[tbh]
\caption{\label{tab:Network-stats}Comparison between power grid models and
synthetic networks}

\centering
\hspace*{-0.4in}{\footnotesize }\begin{tabular}{>{\raggedright}m{0.51in}|>{\raggedright}m{0.36in}>{\raggedright}m{0.36in}>{\raggedright}m{0.36in}>{\raggedright}m{0.36in}|>{\raggedright}m{0.36in}>{\raggedright}m{0.36in}>{\raggedright}m{0.36in}>{\raggedright}m{0.36in}|>{\raggedright}m{0.36in}>{\raggedright}m{0.35in}>{\raggedright}m{0.36in}>{\raggedright}m{0.36in}|>{\raggedright}m{0.36in}>{\raggedright}m{0.36in}>{\raggedright}m{0.36in}>{\raggedright}m{0.36in}}
\hline 
{\scriptsize Network} & \multicolumn{4}{c|}{{\scriptsize Power network}} & \multicolumn{4}{c|}{{\scriptsize Random}} & \multicolumn{4}{c|}{{\scriptsize Small world}} & \multicolumn{4}{c}{{\scriptsize Preferential attachment}}\tabularnewline
 & {\scriptsize IEEE}{\scriptsize \par}

{\scriptsize 300} & {\scriptsize EI} & {\scriptsize WI} & {\scriptsize TI} &  &  &  &  & {\scriptsize $p=0.08$} & {\scriptsize $p=0.0882$} & {\scriptsize $p=0.18$} & {\scriptsize $p=0.22$} &  &  &  & \tabularnewline
\hline
{\scriptsize Nodes} & {\scriptsize 300} & {\scriptsize 41228} & {\scriptsize 11432} & {\scriptsize 4513} & {\scriptsize 300} & {\scriptsize 41228} & {\scriptsize 11432} & {\scriptsize 4513} & {\scriptsize 300} & {\scriptsize 41228} & {\scriptsize 11432} & {\scriptsize 4513} & {\scriptsize 300} & {\scriptsize 41228} & {\scriptsize 11432} & {\scriptsize 4513}\tabularnewline
{\scriptsize Links} & {\scriptsize 409} & {\scriptsize 52075} & {\scriptsize 13734} & {\scriptsize 5532} & {\scriptsize 409} & {\scriptsize 52075} & {\scriptsize 13734} & {\scriptsize 5532} & {\scriptsize 402} & {\scriptsize 52209} & {\scriptsize 13737} & {\scriptsize 5527} & {\scriptsize 409} & {\scriptsize 52032} & {\scriptsize 13719} & {\scriptsize 5521}\tabularnewline
{\scriptsize $<k>$} & {\scriptsize 2.73} & {\scriptsize 2.53} & {\scriptsize 2.40} & {\scriptsize 2.45} & {\scriptsize 2.73} & {\scriptsize 2.53} & {\scriptsize 2.40} & {\scriptsize 2.45} & {\scriptsize 2.68} & {\scriptsize 2.52} & {\scriptsize 2.40} & {\scriptsize 2.45} & {\scriptsize 2.73} & {\scriptsize 2.52} & {\scriptsize 2.40} & {\scriptsize 2.45}\tabularnewline
{\scriptsize $\max(k)$} & {\scriptsize 11} & {\scriptsize 29} & {\scriptsize 22} & {\scriptsize 18} & {\scriptsize 7} & {\scriptsize 12} & {\scriptsize 11} & {\scriptsize 11} & {\scriptsize 6} & {\scriptsize 6} & {\scriptsize 6} & {\scriptsize 6} & {\scriptsize 32} & {\scriptsize 419} & {\scriptsize 185} & {\scriptsize 125}\tabularnewline
{\scriptsize $C$} & {\scriptsize 0.11} & {\scriptsize 0.068} & {\scriptsize 0.073} & {\scriptsize 0.031} & {\scriptsize 0.008} & {\scriptsize 0.00008} & {\scriptsize 0.00015} & {\scriptsize 0.0001} & {\scriptsize 0.26} & {\scriptsize 0.26} & {\scriptsize 0.16} & {\scriptsize 0.15} & {\scriptsize 0.008} & {\scriptsize 0.00054} & {\scriptsize 0.002} & {\scriptsize 0.004}\tabularnewline
{\scriptsize $L$} & {\scriptsize 9.9} & {\scriptsize 31.9} & {\scriptsize 26.1} & {\scriptsize 14.9} & {\scriptsize 5.7} & {\scriptsize 11.2} & {\scriptsize 10.3} & {\scriptsize 9.02} & {\scriptsize 9.6} & {\scriptsize 34.2} & {\scriptsize 22.9} & {\scriptsize 16.9} & {\scriptsize 4.4} & {\scriptsize 7.1} & {\scriptsize 6.8} & {\scriptsize 6.1}\tabularnewline
{\scriptsize $d_{\max}$} & {\scriptsize 24} & {\scriptsize 94} & {\scriptsize 61} & {\scriptsize 37} & {\scriptsize 12} & {\scriptsize 27} & {\scriptsize 24} & {\scriptsize 20} & {\scriptsize 24} & {\scriptsize 101} & {\scriptsize 59} & {\scriptsize 41} & {\scriptsize 9} & {\scriptsize 17} & {\scriptsize 17} & {\scriptsize 15}\tabularnewline
{\scriptsize $r$} & {\scriptsize -0.22} & {\scriptsize -0.10} & {\scriptsize -0.08} & {\scriptsize -0.09} & {\scriptsize 0.044} & {\scriptsize 0.004} & {\scriptsize -0.01} & {\scriptsize -0.03} & {\scriptsize 0.034} & {\scriptsize 0.12} & {\scriptsize 0.052} & {\scriptsize 0.03} & {\scriptsize -0.19} & {\scriptsize -0.03} & {\scriptsize -0.05} & {\scriptsize -0.09}\tabularnewline
\hline
{\scriptsize Hyp. 1} & {\scriptsize -} & {\scriptsize -} & {\scriptsize -} & {\scriptsize -} & {\scriptsize not reject} & {\scriptsize reject{*}{*}} & {\scriptsize reject{*}{*}} & {\scriptsize reject{*}{*}} & {\scriptsize reject{*}{*}} & {\scriptsize reject{*}{*}} & {\scriptsize reject{*}{*}} & {\scriptsize reject{*}{*}} & {\scriptsize reject{*}{*}} & {\scriptsize reject{*}{*}} & {\scriptsize reject{*}{*}} & {\scriptsize reject{*}{*}}\tabularnewline
\hline
{\scriptsize Hyp. 2} & {\scriptsize mar-}\\
{\scriptsize ginal} & {\scriptsize reject{*}{*}} & {\scriptsize reject{*}{*}} & {\scriptsize reject{*}} & {\scriptsize reject{*}} & {\scriptsize reject{*}{*}} & {\scriptsize reject{*}{*}} & {\scriptsize reject{*}{*}} & {\scriptsize reject{*}{*}} & {\scriptsize reject{*}} & {\scriptsize reject{*}{*}} & {\scriptsize reject{*}{*}} & {\scriptsize not reject} & {\scriptsize not reject} & {\scriptsize not reject} & {\scriptsize not reject}\tabularnewline
{\scriptsize est. of $\alpha$} & {\scriptsize 3.5} & {\scriptsize 3.35} & {\scriptsize 3.33} & {\scriptsize 3.44} & {\scriptsize 3.5} & {\scriptsize 3.5} & {\scriptsize 3.5} & {\scriptsize 3.5} & {\scriptsize 3.5} & {\scriptsize 3.29} & {\scriptsize 3.42} & {\scriptsize 3.25} & {\scriptsize 2.49} & {\scriptsize 2.8} & {\scriptsize 2.92} & {\scriptsize 2.55}\tabularnewline
\hline
\end{tabular}{\footnotesize \par}

{\scriptsize {*} Significant at the 0.01 confidence level.}{\scriptsize \par}

{\scriptsize {*}{*} Significant at the 0.001 confidence level.}%
\end{table*}

\begin{figure}[tbh]
\includegraphics[width=0.7\columnwidth]{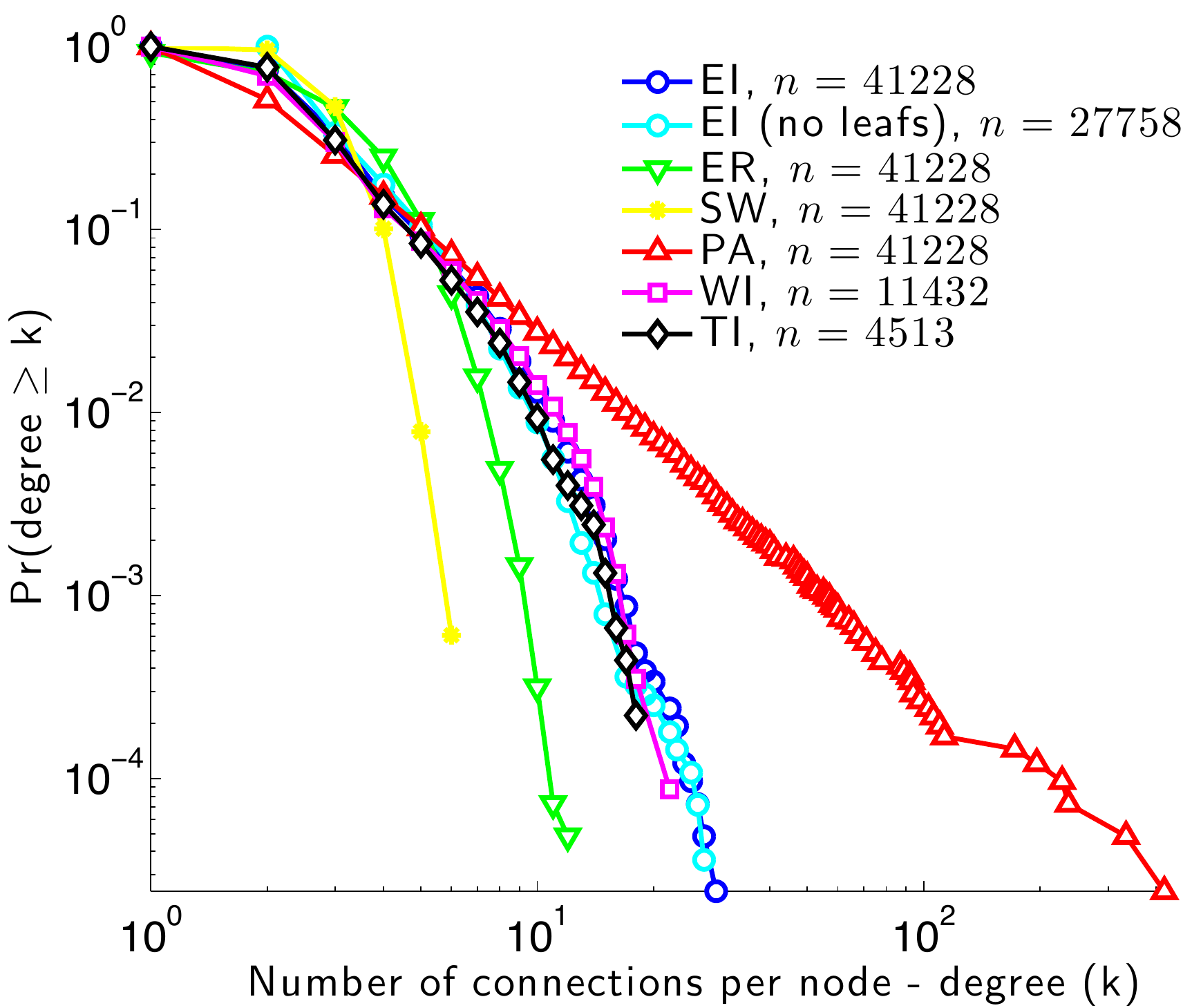}\caption{\label{fig:degree-distributions-41228}Cumulative degree distributions
of the Eastern (EI), Western (WI) and Texas (TI) interconnections,
along with degree distributions for synthetic networks---random (Erd\H{o}s\textendash{}Rényi,
ER), small-world (SW) and preferential-attachment (PA)---sized to
match EI: $|G_{EI}|=\{41228,52075\}$. Power networks have a heavy-tailed,
but not power-law (scale-free) degree distribution.}
\end{figure}

\begin{figure}[tbh]
\includegraphics[width=0.7\columnwidth]{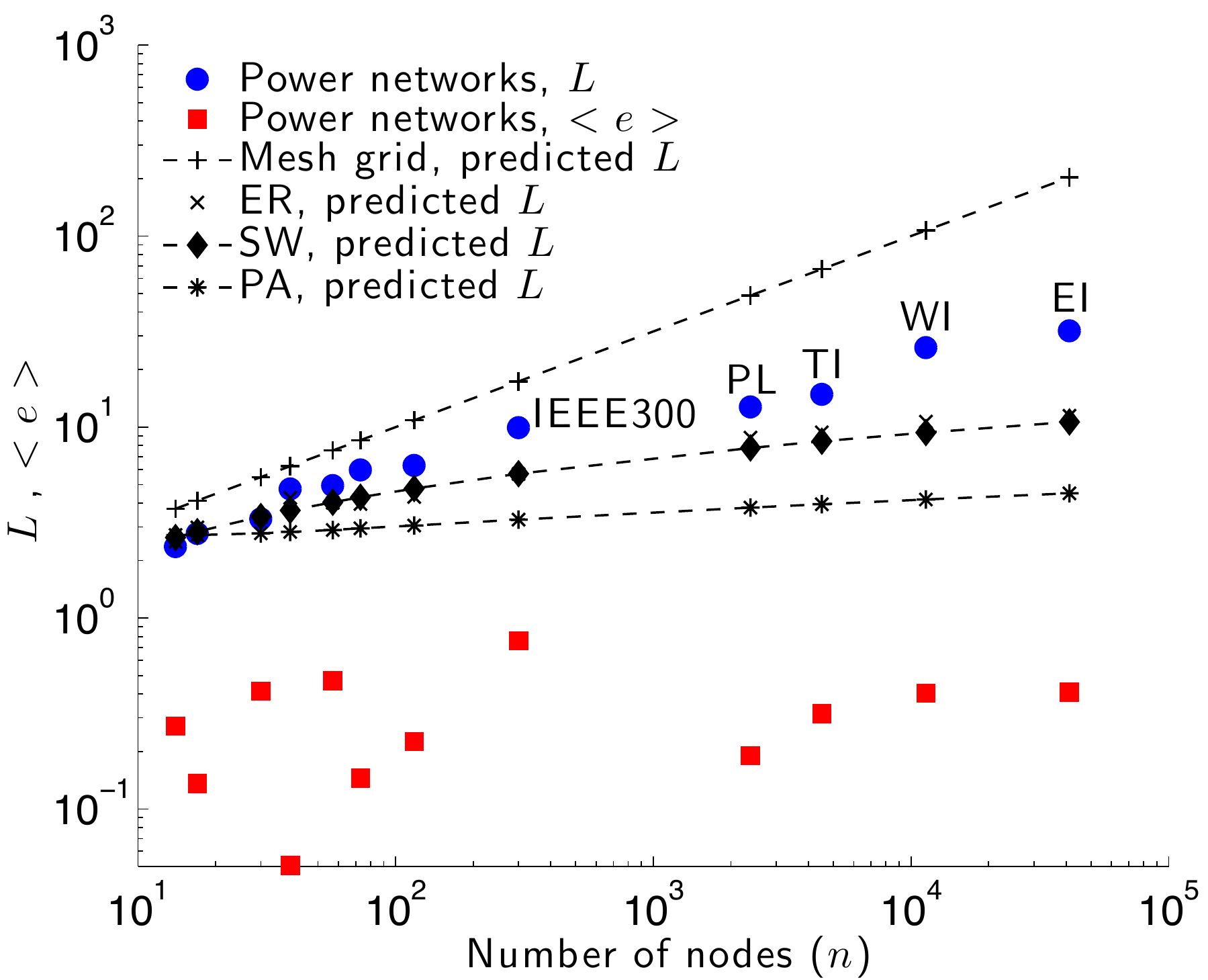}\caption{\label{fig:lbar_and_ebar}Characteristic path length ($L$) and average
electrical distance ($\left\langle e\right\rangle $) of the power
networks as a function of graph size, $n$. $L$ increases faster
than $\ln n$ in the power networks, which indicates that the small-world
model does not fit particularly well. PL is the test system ``case2383wp''
from \cite{Zimmerman:2011}. The smaller networks are standard IEEE
test cases, also available from \cite{Zimmerman:2011}.}
\end{figure}

Tests of hypothesis~1 indicate that the degree distributions of the
power networks do not match those of the synthetic networks in all
but one case: the IEEE 300 network compared to a random graph. In
Fig. \ref{fig:degree-distributions-41228}, which shows the degree
distributions for the large graphs, it is clear that high-degree nodes
are more frequent than would be found in random graphs or small-world
networks, but far less frequent than would be found in a similarly
sized scale-free network. Hypothesis 1 can be rejected with high confidence. 

The tests of hypothesis 2 show, with high confidence, that none of
the North American power network degree distributions follow a power-law
($P<0.001$). The power-law exponents found ($\alpha=3.33$ to $\alpha=3.5$)
are steeper than what would generally be considered notable power-law
distributions \cite{Clauset:2009}. As expected the scale-free network
fits well with a power-law degree distribution. These results lead
us to conclude, with confidence, that power networks are not scale-free
in topological structure.

The conjecture that power networks are small-world in structure requires
some additional analysis. While the clustering coefficients of the
power grids are less that those of the small-world graphs, $C$ is
an order of magnitude higher in power grids than in random networks
(table \ref{tab:Network-stats}). To test whether the small-world
model is a good model for power grids, we introduce a third hypothesis.
Hypothesis 3 is that the characteristic path lengths of power grids
increase with an upper bound of $L\leq\ln n$, which is argued in
\cite{Watts:1998} and \cite{Boccaletti:2006} to be the limit for
distances in small-world networks. Figure \ref{fig:lbar_and_ebar}
compares the characteristic path lengths of several power networks
with predicted values for $L$ from the small-world, scale-free, and
random graph models. Path lengths in large power grids increase substantially
faster than $\ln n$, which means that power grids fall somewhere
between a regular grid, in which path lengths scale linearly with
$n$ or $\sqrt{n}$, and a small-world network. Because $L>\ln n$
for all of the power networks of substantial size ($n>30$) hypothesis
3 can be safely rejected. Therefore, methods that are based on the
assumption that power grids belong to the small-world family of networks
(e.g., \cite{Fu:2010}) may be misleading. 

A number of other results in table \ref{tab:Network-stats} and Fig.
\ref{fig:degree-distributions-41228} are notable. The small world
networks had much higher clustering than the power networks, largely
because the re-wiring probabilities ($p$) were chosen to keep distances
similar to those found in the power networks, and were thus very small.
That the power networks had a fairly high degree of clustering may
be the result of geographic constraints that make very long-distance
connections impractical in most cases. Also we find that power networks
have a small, negative degree assortativity. In contrast, the small-world
model shows a positive degree correlation, and, as expected, the preferential-attachment
and random graph models show nearly zero assortativity. The negative
assortativity in the power networks was found to result from substation
buses with a large number of radial connections (typically distribution
feeders). If the degree-1 ``leaf nodes'' are recursively removed
the assortativity becomes very nearly zero. Upon recalculating the
assortativity of EI, WI and TI after removing the leaf nodes, we find
the following changes in assortativity: EI, from $r=-0.10$ to $-0.02$;
WI, from $r=-0.08$ to $-0.01$; and TI, from $r=-0.09$ to $-0.02$.

\section{The electrical structure of power grids\label{sec:electrical}}

While a thorough understanding of topological structure can be useful
for some problems, such as producing synthetic power grids \cite{Wang:2010},
data about topological structure are not sufficient to describe the
performance of power networks \cite{Hines:2010b}. The connections
between components in a power grid depend not only on its topology,
but also on the the physical properties that govern voltages and currents. 

One way to study connectivity between components in an electrical
system is to look at the properties of sensitivity matrices. Sensitivity
matrices for power systems, such as power transfer distribution factor
matrices \cite{Bialek:1997}, measure the amount of influence that
one component has on another. The complement of a sensitivity matrix
is a distance matrix, which measures zero for component pairs that
are perfectly connected, and large number for component pairs that
have very little influence on one another. To our knowledge, the earliest
work on electrical distances in power systems is in \cite{Lagonotte:1989},
in which the authors propose a distance metric based on voltage magnitude
sensitivities and use this metric to divide a network into voltage
control zones. They showed that the logarithmic voltage magnitude
sensitivity in a power grid can qualify as a formal distance metric
(see Appendix), under some conditions. Electrical distance measures
have been applied to a number of reliability and economic power system
problems \cite{Lu:1995,Hang:2000,Zhong:2004}. 

The concept of ``Resistance Distance,'' as proposed in \cite{Klein:1993},
provides another method for computing electrical distances. Resistance
distances give the effective resistance between points in a network
of resistors. Here we extend the idea of resistance distance to represent
the marginal effect of an active power transaction between buses $a$
and $b$ on voltage phase angle differences. Some transactions in
a power system, such as those that occur over short, low impedance
lines, have relatively minor effects on a network. Other transactions
occur across high impedance paths, and thus result in large phase
angle changes in the network. Large phase angle separations can expose
instabilities in the system. Reference \cite{Dobson:2010b} proposes
the use of voltage phase angle differences between areas as a measure
of stress in power networks. With this in mind we propose a distance
metric based on the sensitivity between active power transfers and
nodal phase angle differences.

To do so we start with a network of resistors with current injections
at each node. Such a network can be described by a conductance matrix
$ $$\mathbf{G}$, such that the current injection at node $a$ is:
\begin{equation}
I_{a}=\sum_{b=1}^{n}g_{ab}V_{b}\label{eq:currentinjection}
\end{equation}
When there are no connections to ground in the circuit, $\mathbf{G}$
has rank $n-1$, unless a voltage reference is specified. $\mathbf{G}$,
thus defined acts as a Laplacian matrix for the resistive network.
To get around the singularity of $\mathbf{G}$, we let node $r$ be
a voltage reference node, with $V_{r}\triangleq0$. The sub-matrix
of $\mathbf{G}$ associated with the non-reference nodes ($\bar{r}$)
is full-rank, therefore we can compute a sub-matrix \textbf{$\mathbf{G}_{\bar{r}\bar{r}}^{-1}$},
such that: 
\begin{equation}
\mathbf{V}_{\bar{r}}=\mathbf{G}_{\bar{r}\bar{r}}^{-1}\mathbf{I}_{\bar{r}}.\label{eq:Vr}
\end{equation}
If the diagonal element of $\mathbf{G}_{\bar{r}\bar{r}}^{-1}$ associated
with node $a$ is $g_{a,a}^{-1}$, $g_{a,a}^{-1}$ indicates the change
in voltage between nodes $a$ and $r$ as a result of a current injection
at node $a$, which must be withdrawn from node $r$. This resistance
distance gives the sensitivity between current injections and voltage
differences. Therefore the resistance distance between $a$ and $r$
is equal to $g_{a,a}^{-1},\forall a$. To measure the distance between
a pair of nodes $a$ and $b$, where $a\neq b\neq r$, one can evaluate:
\begin{equation}
e(a,b)=g_{a,a}^{-1}+g_{b,b}^{-1}-g_{a,b}^{-1}-g_{b,a}^{-1},\label{eq:e(a,b)}
\end{equation}
which gives the voltage difference between $a$ and $b$ after injecting
1 A at $a$ and withdrawing 1 A from $b$. In matrix form, where $\gamma\triangleq\mbox{diag}(\mathbf{G}_{\bar{r}\bar{r}}^{-1})$,
the calculation proceeds efficiently as follows:
\begin{eqnarray}
\mathbf{E}_{\bar{r}\bar{r}} & = & \mathbf{1}\gamma^{T}+\gamma\mathbf{1}^{T}-\mathbf{G}_{\bar{r}\bar{r}}^{-1}-[\mathbf{G}_{\bar{r}\bar{r}}^{-1}]^{T}\label{eq:E_notr_notr}\\
\mathbf{E}_{r,\bar{r}} & = & \gamma^{T}\\
\mathbf{E}_{\bar{r},r} & = & \gamma
\end{eqnarray}
where $\mathbf{E}_{\bar{r},r}$ refers to vector of non-reference
elements in the reference column of the distance matrix. Resistance
distance, thus defined, is proven to have the properties of a metric
space in \cite{Klein:1993}. 

To obtain sensitivities between power injections and phase angles,
this definition requires some modification. To produce a real-valued
distance metric, we start with the top half of the power flow Jacobian
matrices:
\begin{equation}
\Delta\mathbf{P}=\left[\frac{\partial P}{\partial\theta}\right]\Delta\mathbf{\theta}+\left[\frac{\partial P}{\partial|V|}\right]\Delta|V|\label{eq:dp}
\end{equation}
If we assume that voltages are held constant ($\Delta|V|=0$), the
matrix $[\partial P/\partial\theta]$ can be used to form a distance
matrix. Under the dc power flow assumptions (nominal voltages, no
resistive losses and small angle differences), $[\partial P/\partial\theta]$
is a Laplacian matrix that is analogous to $\mathbf{G}$. Even when
these assumptions are relaxed, $[\partial P/\partial\theta]$ has
most of the properties of a Laplacian. Applying (\ref{eq:e(a,b)})
with $\mathbf{G}=[\partial P/\partial\theta]$ (with or without the
dc approximations) results in a distance matrix ($\mathbf{E}$) that
measures the incremental change in phase angle difference between
nodes $a$ and $b$ ($\theta_{a}-\theta_{b}$) given an incremental
average power transaction between nodes $a$ and $b$, assuming that
voltage magnitudes are held constant. As in \cite{Lagonotte:1989},
we empirically found that $\mathbf{E}$, thus defined, satisfies the
properties of a distance matrix so long as all series branch reactances
are non-negative, as would be the case for nodes connected by a series
capacitor (see Appendix).

\begin{figure}[tbh]
\includegraphics[width=0.7\columnwidth]{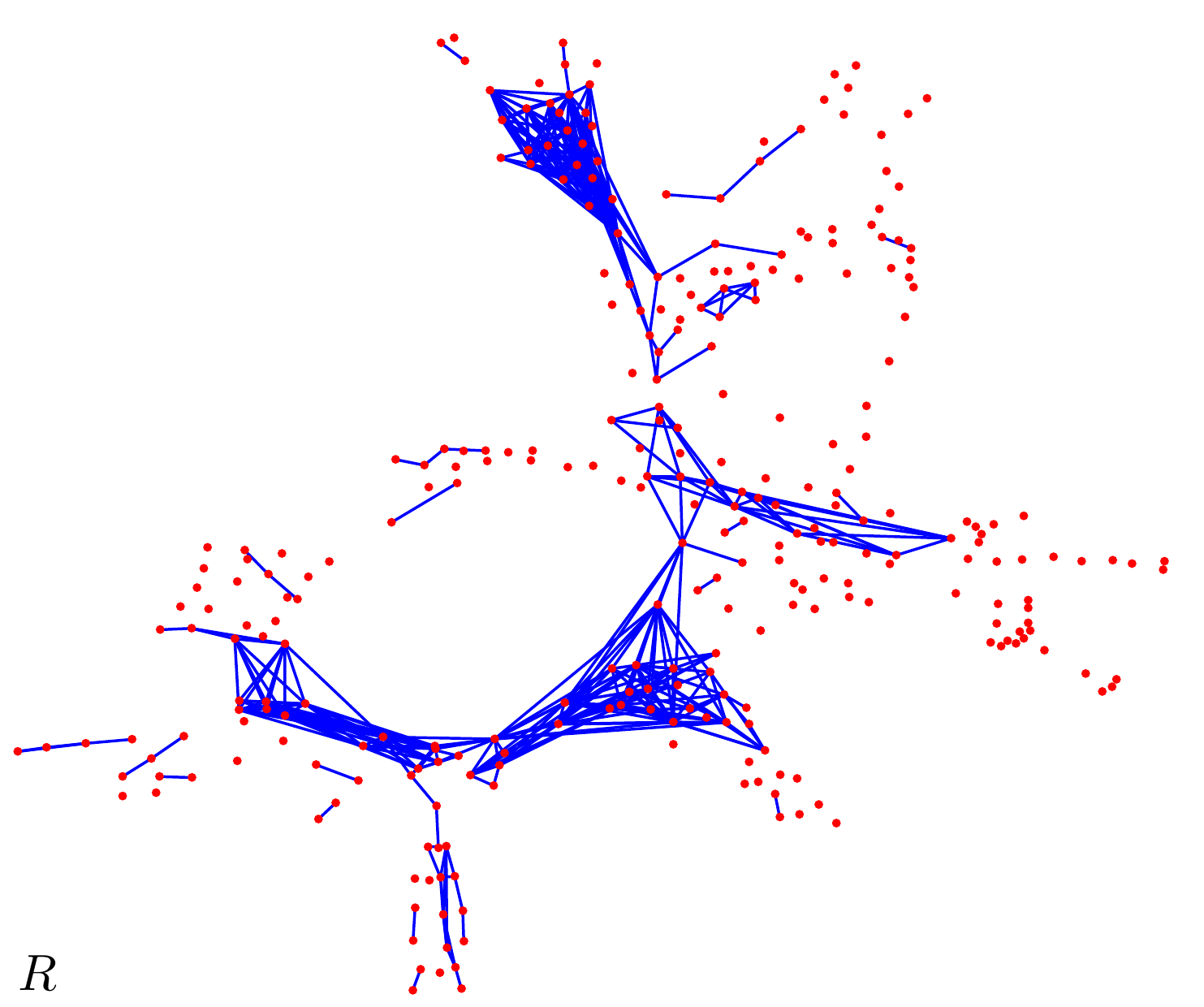}

\includegraphics[width=0.7\columnwidth]{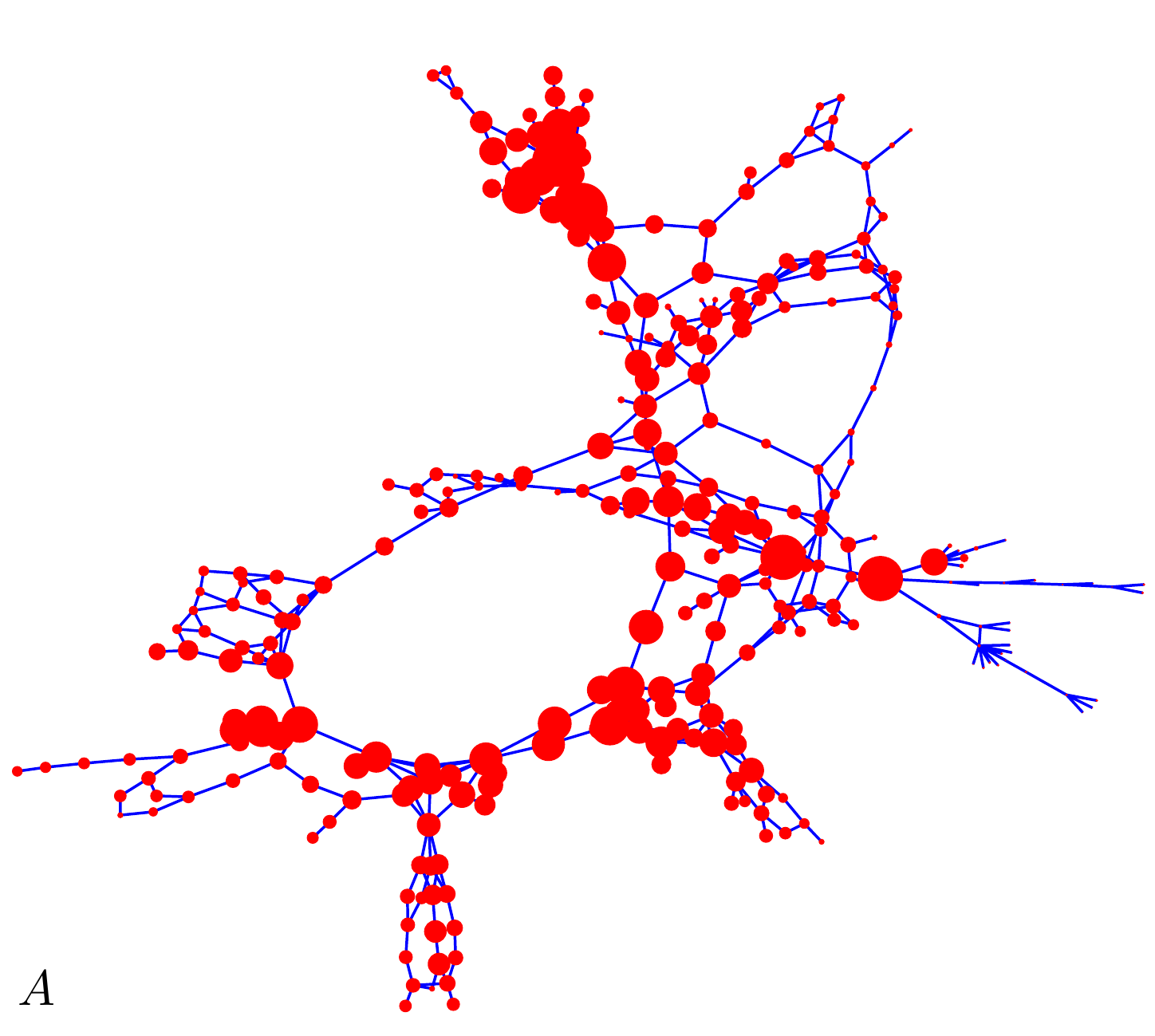}

\caption{\label{fig:electric-topology}The upper panel shows a graphical representation
of the electrical topology of the IEEE 300 bus system ($\mathbf{R}$),
formed by replacing the 409 transmission branches with 409 shortest
distance electrical links (the 409 smallest $e_{ij}$ such that $i>j$).
The lower panel shows the topology ($\mathbf{A}$) of the 300 bus
network with nodes drawn with sizes proportional to their electrical
centrality (\ref{eq:electricalcentrality}).}
\end{figure}

Using this metric we can define a measure that is roughly analogous
to node degree, but for a fully connected network with continuous
weights for each node pair. To do so we measure the average distance
from each node $a$ to other nodes in the system:
\begin{equation}
<e_{a}>=\sum_{b=1}^{n}\frac{e_{ab}}{n-1}\label{eq:electricaldistance}
\end{equation}
and invert it to obtain a measure of electrical centrality (modified
from \cite{Hines:2008}):
\begin{equation}
c_{a}=\frac{1}{<e_{a}>}.\label{eq:electricalcentrality}
\end{equation}
The lower panel of Fig. \ref{fig:electric-topology} illustrates the
electrical centrality of buses in the IEEE 300 bus system. In this
system there are a group of buses with very high electrical connectivity,
whereas the majority of buses have minimal centrality.

\begin{figure}[tbh]
\includegraphics[width=0.7\columnwidth]{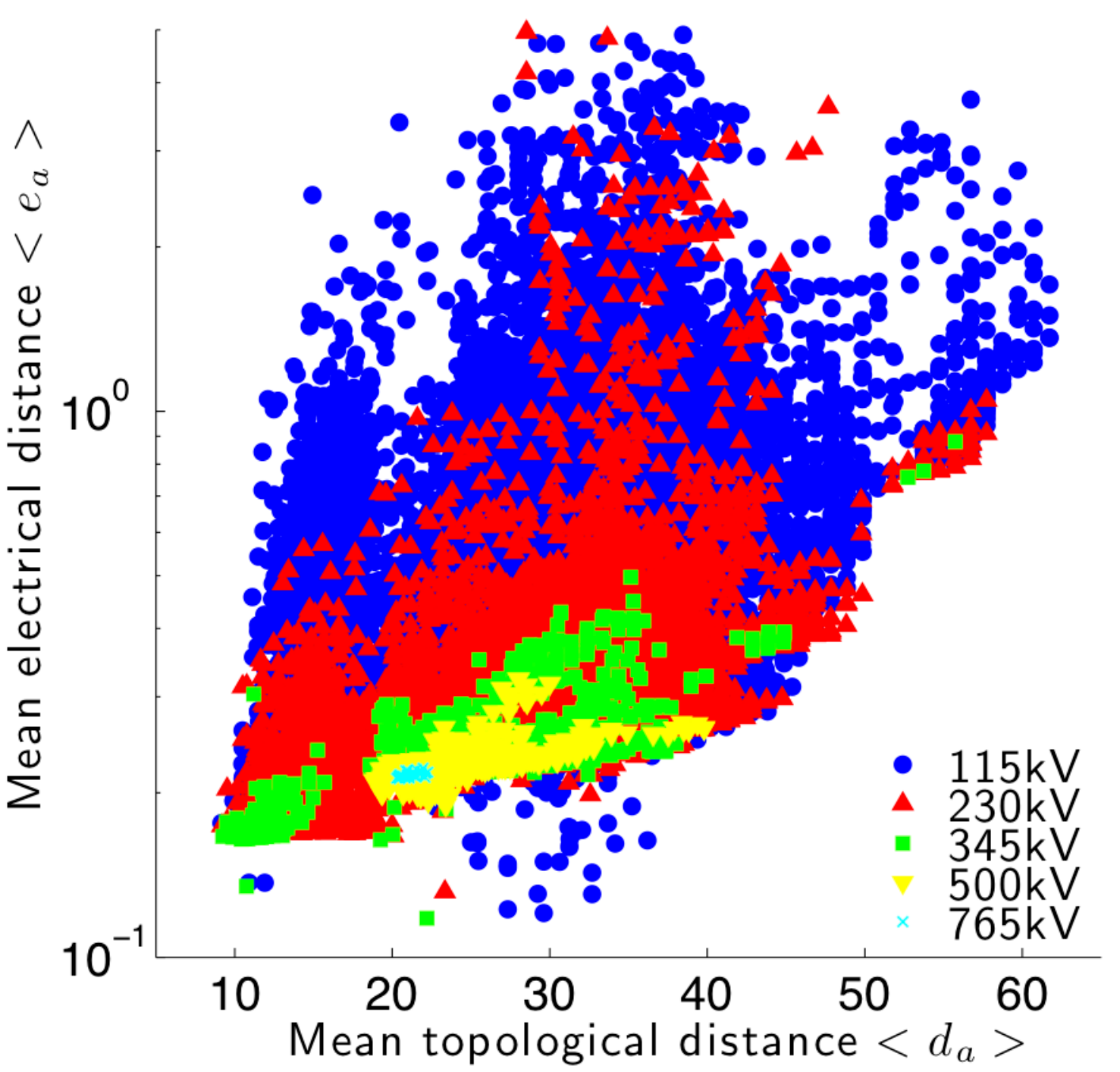}\caption{\label{fig:di_and_ei_scatter}The mean topological and electrical
(\ref{eq:electricaldistance}) distances of buses, at different voltage
levels, within the three North American power networks (EI, WI and
TI). The average correlation between topological distance and electrical
distance is low ($\rho=0.24$), however the correlation at the higher
voltage levels ($\rho_{765}=0.43$, $\rho_{500}=0.65$) is higher.}
\end{figure}

Electrical distances are notably different than topological distances.
Figure \ref{fig:di_and_ei_scatter} shows that there is only a very
weak correlation between electrical and topological distances ($\rho=0.24$
for $\left\langle e_{a}\right\rangle $ and $\left\langle d_{a}\right\rangle $).
Many node pairs that are close topologically are distant in terms
of electrical distance. It is notable that the correlation is stronger
at the higher (500kV and 765kV) voltage levels where there is less
diversity in branch impedances. 

\begin{figure}[tbh]
\includegraphics[width=0.7\columnwidth]{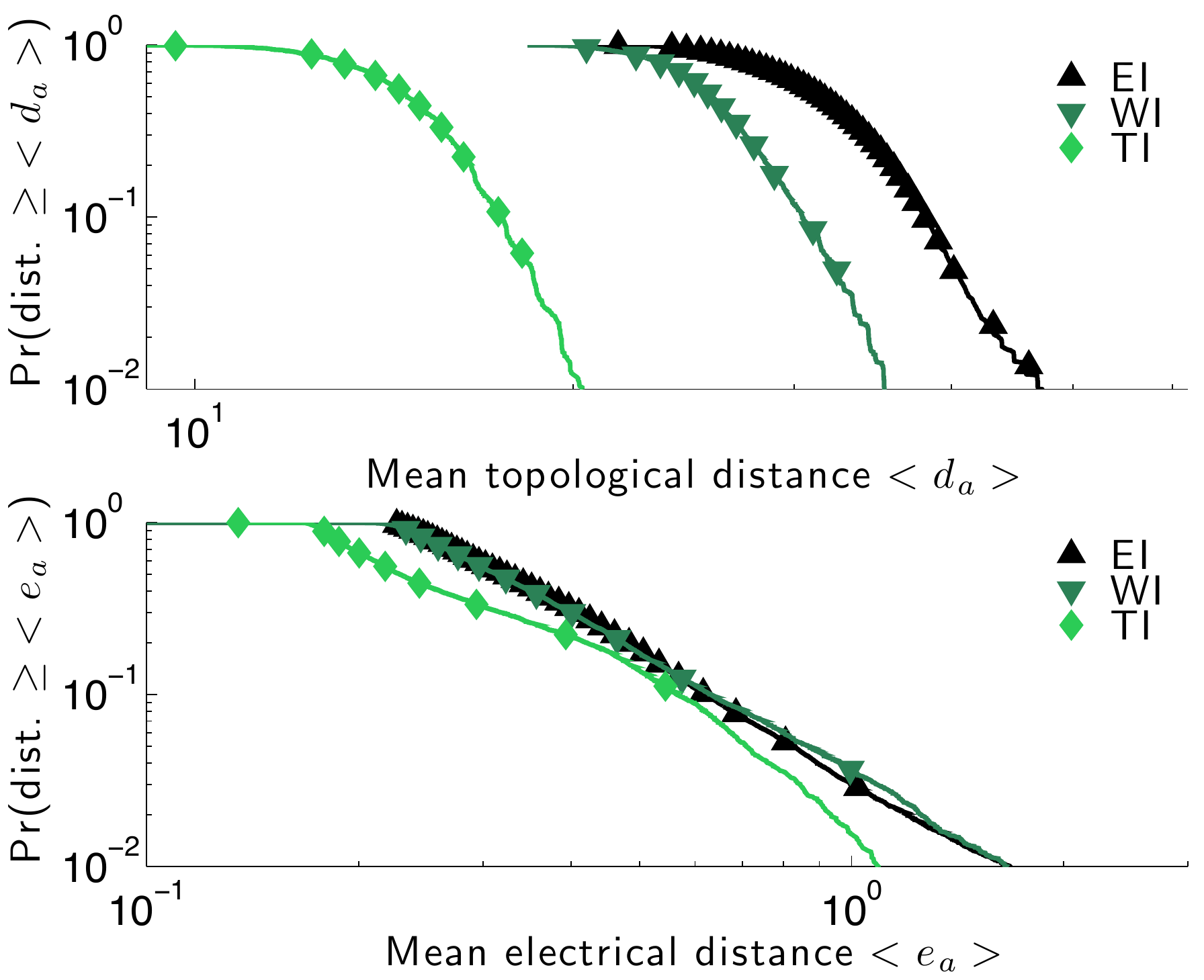}

\caption{\label{F:DEdegreedist}Complementary cumulative probability distributions
for electrical and topological distances ($d_{ab}$ and $e_{ab}$)
for the US Eastern, Western and Texas Interconnections. For the three
distributions of topological distances, hypothesis 2 (see Section
\ref{sub:TopologicalResults}) is rejected ($p<0.001$). For the two
larger distributions of electrical distances (EI and WI), hypothesis
2 is not rejected and the estimates of $\alpha$ are: $\alpha_{EI}=3.49$
and $\alpha_{WI}=3.36$. For the distribution of electrical distances
in TI, hypothesis 2 is rejected ($p<0.001$).}
\end{figure}

Figure \ref{F:DEdegreedist} compares the distribution of electrical
and topological distances obtained from the EI, WI and TI models.
The data indicate that topological distance distributions ($\mathbf{D}$)
have exponential tails, whereas the electrical distance distributions
($\mathbf{E}$) have power-law tails. 

\subsection{Representing electrical distances as an unweighted graph}

The electrical distance matrix $\mathbf{E}$ describes the amount
of connectivity between all node pairs in the system. Because Kirchhoff's
and Ohm's laws produce comprehensive connectivity among all nodes
in the system, the graph is weighted and fully connected (i.e., the
network is an undirected graph with $n(n-1)$ weighted links). In
order to compare this network with undirected, unweighted networks,
we used the electrical distance matrix to produce an unweighted graph
representation of the electrical structure of a power network. To
do so, we kept the original $n$ nodes, but replaced the $m$ links
with the $m$ smallest entries in the upper (or lower since $\mathbf{E}$
is symmetric) triangle of $\mathbf{E}$. Thus we created a graph of
size $\{n,m\}$ with links representing strong electrical connections
rather than direct physical connections. The adjacency matrix of this
new graph ($\mathbf{R}$) is defined as follows: 
\begin{equation}
\mathbf{R}:\,\begin{cases}
r_{ab}=1 & \forall e_{ab}<t\\
r_{ab}=0 & \forall e_{ab}\geqslant t
\end{cases}\label{eq:r_matrix}
\end{equation}
where $t$ is a threshold adjusted to produce exactly $m$ links in
the network. The upper panel of Fig. \ref{fig:electric-topology}
shows the resulting electrical structure graph ($\mathbf{R}$) for
the IEEE 300 bus test case. Comparison of the topological graph ($\mathbf{A}$
in Fig. \ref{fig:electric-topology}) to the electrical one ($\mathbf{R}$
in Fig. \ref{fig:electric-topology}) shows a stark contrast between
the electrical and topological structure of the test system. A similar
structural difference was found by comparing the topology and electrical
structure of North American systems. Table \ref{tab:Redmetrics} summarizes
the calculated metrics for the Eastern Interconnection $\mathbf{R}$
matrix. Electrically speaking, a few nodes have a very high connectivity,
whereas the vast majority of vertices in $\mathbf{R}$ have no connections.
In this way, power systems share some properties with the scale-free
network model; a few nodes have disproportionate influence on a large
portion of the network. However, this is not to say that power systems
will have all of the properties of scale-free networks (such as being
highly vulnerable to directed attacks), since the response of power
systems to disturbances depends on a wide variety of factors that
are not captured in electrical distances.

The clustering coefficient of $\mathbf{R}$ is also high relative
to the value obtained from the equivalent topological network. Both
path lengths ($L$) and diameter ($d_{\max}$) are smaller in the
electrical graph, indicating strong electrical connectedness within
the core. The WI and TI electrical networks are highly disassortative
($r<0)$, whereas the EI system is electrically assortative ($r=0.33$). 

\begin{table*}[tbh]
\caption{\label{tab:Redmetrics}Topologically equivalent metrics for the reduced
electrical distance matrices ($\mathbf{R}$).{*}}

\centering

{\footnotesize }%
\begin{tabular}{>{\centering}p{0.8in}>{\centering}p{0.8in}>{\centering}p{0.8in}>{\centering}p{0.8in}>{\centering}p{0.8in}>{\centering}p{0.8in}>{\centering}p{0.8in}}
\hline 
{\footnotesize Network} & {\footnotesize EI ($\mathbf{R}$)} & {\footnotesize EI ($\mathbf{A}$) } & {\footnotesize WI ($\mathbf{R}$)} & {\footnotesize WI ($\mathbf{A}$) } & {\footnotesize TI ($\mathbf{R}$)} & {\footnotesize TI ($\mathbf{A}$) }\tabularnewline
\hline 
{\footnotesize Nodes{*}{*}} & {\footnotesize 41228 (56)} & {\footnotesize 41228} & {\footnotesize 11432 (8017)} & {\footnotesize 11432} & {\footnotesize 4513 (2769)} & {\footnotesize 4513}\tabularnewline
{\footnotesize Links} & {\footnotesize 52075} & {\footnotesize 52075} & {\footnotesize 13734} & {\footnotesize 13734} & {\footnotesize 5532} & {\footnotesize 5532}\tabularnewline
{\footnotesize $<k>$} & {\footnotesize 2.53} & {\footnotesize 2.53} & {\footnotesize 2.40} & {\footnotesize 2.40} & {\footnotesize 2.45} & {\footnotesize 2.45}\tabularnewline
{\footnotesize $\max(k)$} & {\footnotesize 41171} & {\footnotesize 29} & {\footnotesize 3413} & {\footnotesize 22} & {\footnotesize 762} & {\footnotesize 18}\tabularnewline
{\footnotesize $C$} & {\footnotesize 0.9996} & {\footnotesize 0.068} & {\footnotesize 0.9982} & {\footnotesize 0.073} & {\footnotesize 0.49} & {\footnotesize 0.031}\tabularnewline
{\footnotesize $L$} & {\footnotesize 1.999} & {\footnotesize 31.9} & {\footnotesize 1.998} & {\footnotesize 26.1} & {\footnotesize 2.79} & {\footnotesize 14.9}\tabularnewline
{\footnotesize $d_{max}$} & {\footnotesize 2} & {\footnotesize 94} & {\footnotesize 3} & {\footnotesize 61} & {\footnotesize 13} & {\footnotesize 37}\tabularnewline
{\footnotesize $r$} & {\footnotesize 0.33} & {\footnotesize -0.10} & {\footnotesize -0.96} & {\footnotesize -0.08} & {\footnotesize -0.29} & {\footnotesize -0.09}\tabularnewline
\hline 
\end{tabular}{\footnotesize \par}

{\footnotesize {*} Metrics are calculated for the giant component.}{\footnotesize \par}

{\footnotesize {*}{*} Numbers in parenthesis account for isolated
nodes.}
\end{table*}

\subsection{Electrical distance to load}

\begin{figure}[h]
\includegraphics[width=0.7\columnwidth]{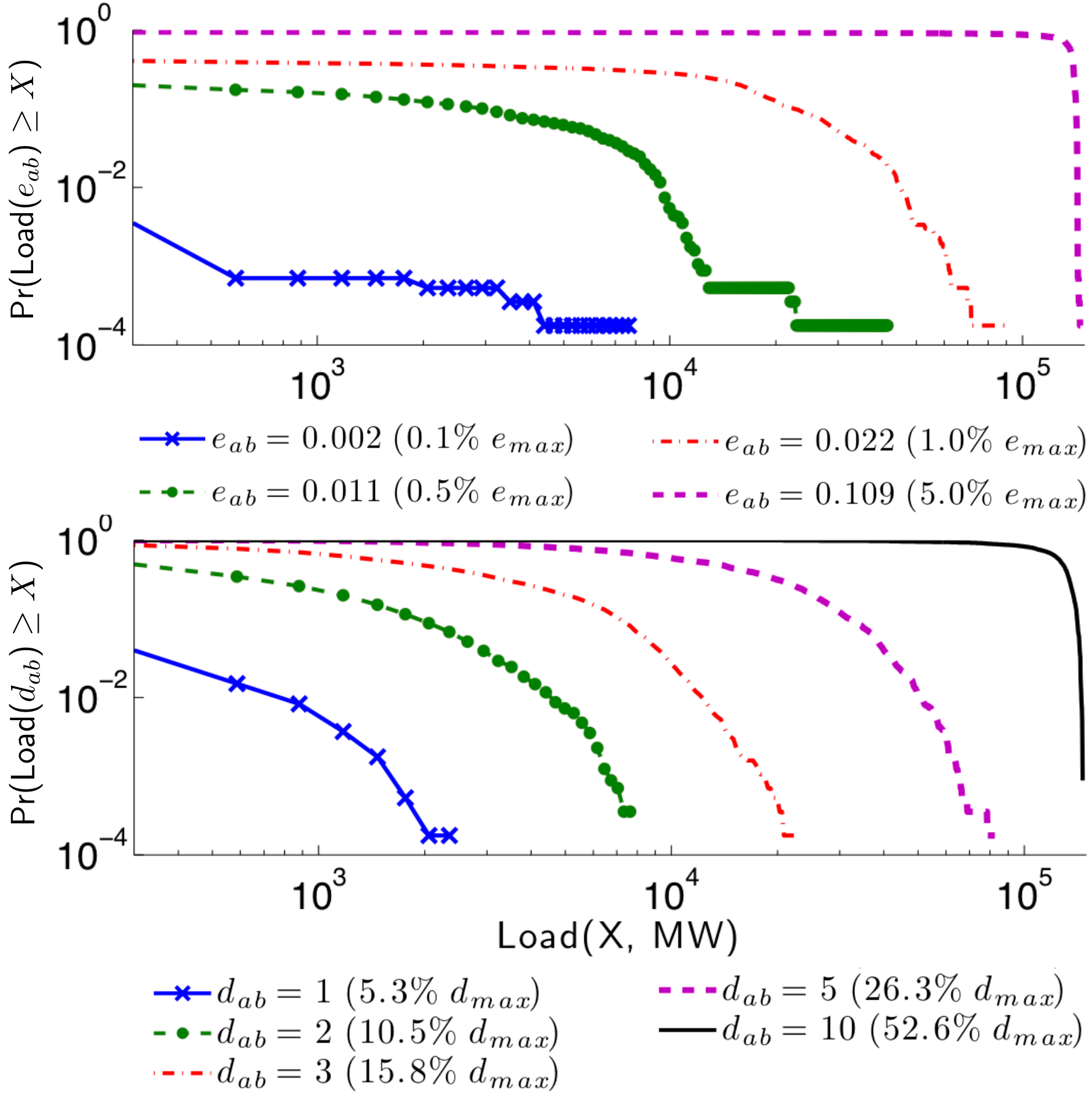}

\caption{\label{fig:load-centrality}Electrical distances to loads for the
PJM portion of the US Eastern Interconnect. The horizontal axis shows
load, in MW. The vertical axis for the top panel shows the probability
of a randomly selected node reaching $X$ MW of load within electrical
distance $e_{ab}$. The lower panel shows the same result for topological
distances. Comparing the purple line in the top figure (5\% of $e_{max}$)
with the blue line in the bottom figure (\textasciitilde{}5\% of $d_{max}$)
we see that it is possible to reach large amounts of load by traversing
over short electrical distances, relative to the topological distances
that one would need to cover to reach the same amount of load.}
\end{figure}

The distance metric described by \eqref{eq:Vr}-\eqref{eq:E_notr_notr}
gives sensitivities between arbitrary node pairs in a power network.
We can refine this metric to focus on electrical distances from an
arbitrary node to load nodes. This \textquotedbl{}electrical distance
to load\textquotedbl{} measure describes the sensitivity of consumption
nodes to perturbations at other locations in the network. In this
section we determine the electrical distance to load of approximately
5600 buses within the US mid-Atlantic (PJM) portion of the EI model.
From the electrical distance matrix, $\mathbf{E}$, we plot (Fig.
\ref{fig:load-centrality}) the amount of the total system load that
is reachable at various electrical distances. Comparing this to a
similar graph for topological distances illustrates the difference
between the two. Topologically there are a few nodes that are within
1 link ($d_{ab}\leq1$, or 5.3\% of the topological diameter, $d_{\max}$)
of about 2 GW of load, which is 1\% of the total system load of 146.8
GW (lower panel of Fig. \ref{fig:load-centrality}). Electrically,
there are a few nodes that are within $e_{ab}\leq0.011$ (0.5\% of
the electrical diameter, $e_{\max}$) of 40 GW (27\%) of load. This
indicates that a small number of nodes have a very high electrical
influence on the system as a whole. Typically, these are high voltage
buses (500kV and larger) that show very high electrical centrality. 

\section{Conclusions\label{sec:conclusions}}

Power grids have been an appealing topic for network analysis in recent
years, but the existing body of literature has produced some mixed
results, particularly related to measuring the structure of power
networks and connections between structure and performance. We have
assessed the topological structure of the North American electric
power network, and have compared it with a structural analysis based
on a measure of ``electrical distance,\textquotedbl{} which measures
network sensitivities. Electrical distance is more closely related
to the physical laws governing behavior in electrical networks than
topological distance or connectivity. In the appendix we show the
conditions under which the proposed distance measure qualifies as
a formal distance metric. Simpler impedance-based measures of electrical
distance used elsewhere in the literature do not always meet the criteria
for a proper distance metric.

Our analysis of the topological structure of the Eastern, Western
and Texas Interconnects clearly suggests an exponential degree distribution
rather than a power-law. We thus reject the conjecture that the observed
power-law distribution in blackout sizes arises due to a power-law
in the grid topology. Further, we observe significant differences
between power-grid topologies and the topology of synthetic small-world
networks. Distances between nodes increase much faster than the log
of network size, and clustering is not as high as is found in true
small-world networks. On a topological basis, we thus conclude that
power grids are neither scale-free nor small-world in topological
structure. 

When we measure network structure based on electrical distance, however,
we find a power-law distribution in electrical distances for all three
Interconnects in the North American power network. It may be that
this power-law distribution contributes to the empirically observed
power-law distribution in large blackout sizes. Future work will seek
to characterize the nature of this relationship. Also, we find that
the correlation between electrical and topological distance is weak,
though it increases for higher-voltage lines. When we look at the
electrical distances between buses and loads in power networks, we
find that a small number of nodes are tightly electrically connected
to large percentages of load in the system, despite being relatively
distant topologically. Since our electrical distance measure captures
network sensitivities rather than physical connectivity, the observed
heterogeneous structure suggests that perturbations (such as those
caused by a volitional attack or changes in generator pricing) may
have non-local effects more often than would be expected from the
grid's topological structure. Our work suggests that future endeavors
to relate structure and performance in electrical networks should
focus on characterizing network sensitivities rather than focusing
narrowly on topological connectivity. Furthermore, this work suggests
that circuit theory can be a useful tool in characterizing the connectivity
and thus performance of complex networks.

\section*{Appendix. Conditions under which electrical distance qualifies as
a formal distance measure}

To qualify as a distance metric, a metric space $d(x,y)$ must map
pairs of elements in a set in a way that describes the dissimilarity
among any element ($x,y,z$), and must have the properties of non-negativity
($d(x,y)\geq0$), symmetry ($d(x,y)=d(y,x)$), identity ($d(x,y)=0\Leftrightarrow x=y$)
and the triangle inequality ($d(x,z)\leq d(x,y)+d(y,z)$) \cite{Rudin:1964}. 

Equations (\ref{eq:e(a,b)})-(\ref{eq:E_notr_notr}) describe our
measure of electrical distance $e(a,b)$, given a sensitivity matrix
$\mathbf{G}$. Here we discuss the conditions under which $e(a,b)$
qualifies as a proper distance metric in some metric space. Following
\cite{Rudin:1964,Klein:1993}, a distance metric $e(x,y)$ on an arbitrary
metric space gives a measure of dissimilarity between arbitrary objects
$x$ and $y$, and satisfies the following properties for all elements
$a,b,c$:
\begin{itemize}
\item Symmetry: $e(a,b)=e(b,a)$
\item Identity: $e(a,b)=0\Longleftrightarrow a=b$
\item Non-Negativity: $e(a,b)\geq0$
\item Triangle Inequality: $e(a,b)+e(b,c)\geq e(a,c)$.
\end{itemize}
The discussion in this appendix will focus on the last of these properties,
the triangle inequality. Lagonotte (\cite{Lagonotte:1989}) found
that an electrical distance measure related to voltages generally,
but not always, obeyed the triangle inequality. We demonstrate a similar
finding here for an analogous distance measure which is based upon
real power sensitivities. Building on \cite{Lagonotte:1989}, we also
characterize the conditions under which the triangle inequality will
fail to hold for this electrical distance measure.

\begin{figure}[h]
\centering

\includegraphics[width=0.7\columnwidth]{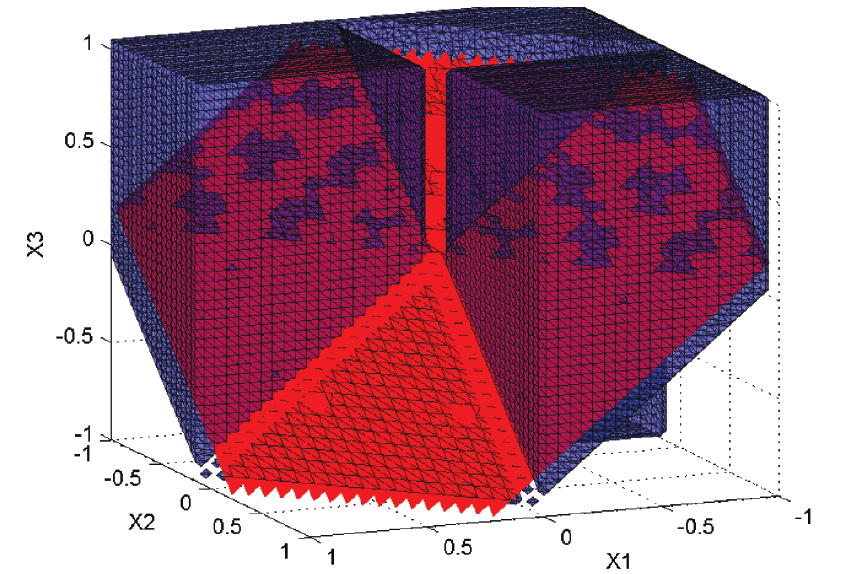}

\includegraphics[width=0.7\columnwidth]{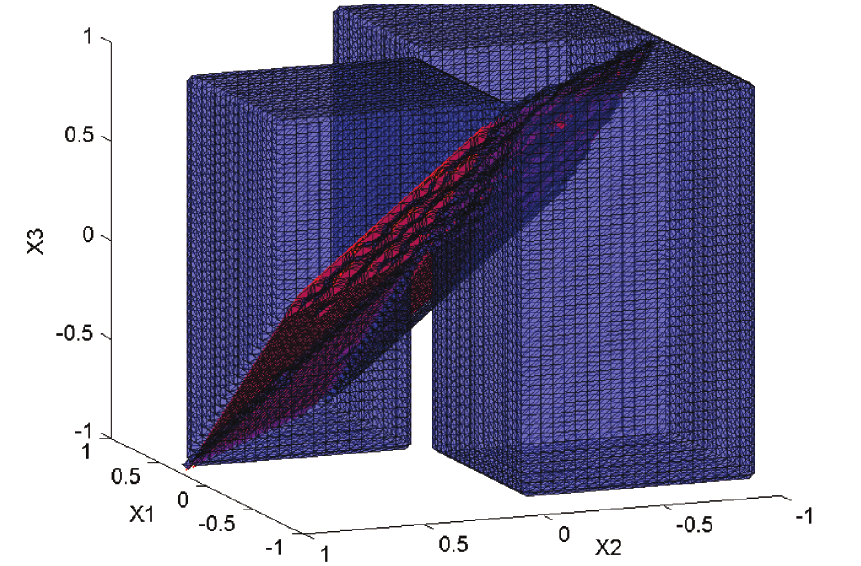}

\caption{\label{fig:triangle-surface}Illustration of the conditions under
which three reactances ($x_{12},x_{23},x_{13})$ connecting a three-bus
network satisfy the triangle inequality. Combinations that result
in a triangle inequality violation are shaded. The two panels show
the conditions from different perspectives. The red plane ($0=x1+x2+x3$)
separates vertically the volumes corresponding to failure and success. }
\end{figure}

Recall, from the discussion surrounding (\ref{eq:E_notr_notr}), that
the sensitivity matrix \textbf{G} is replaced with $[\partial\mathbf{P}/\partial\mathbf{\mathbf{\theta}}]$
from the power flow Jacobian. In the dc power flow model \cite{Stott:2009},
this sensitivity is simply equal to the system susceptance matrix
\textbf{B}. For an arbitrary, connected three-node network, following
(\ref{eq:E_notr_notr}), the electrical distance matrix \textbf{E}
can be written as:
\begin{equation}
\left[\begin{array}{ccc}
0 & \frac{x_{12}(x_{13}+x_{23})}{x_{12}+x_{13}+x_{23}} & \frac{x_{13}(x_{12}+x_{23})}{x_{12}+x_{13}+x_{23}}\\
\frac{x_{12}(x_{13}+x_{23})}{x_{12}+x_{13}+x_{23}} & 0 & \frac{x_{23}(x_{12}+x_{13})}{x_{12}+x_{13}+x_{23}}\\
\frac{x_{13}(x_{12}+x_{23})}{x_{12}+x_{13}+x_{23}} & \frac{x_{23}(x_{12}+x_{13})}{x_{12}+x_{13}+x_{23}} & 0
\end{array}\right]\label{eq:E_3bus}
\end{equation}
Applying the triangle inequality to (\ref{eq:E_3bus}), for each of
the three triplets, leads to the following sufficient condition on
the three reactances:
\begin{eqnarray}
(x_{12}x_{23}\geq0 & \cap & (x_{12}+x_{23}+x_{13})>0)\cup\label{eq:failure}\\
(x_{12}x_{23}<0 & \cap & (x_{12}+x_{23}+x_{13})<0)\nonumber 
\end{eqnarray}
which is clearly satisfied for a strictly positive set of reactances.
Figure \ref{fig:triangle-surface} shows the volume in $x_{12},x_{13},x_{23}$
space where violations of the triangle inequality are observed, for
all combinations of $x_{ab}\in[-1,1]$. Areas of the surface where
\begin{equation}
e(a,b)+e(b,c)-e(a,c)<0\label{eq:tri_ineq}
\end{equation}
were shaded, showing combinations of $x_{12},x_{13},x_{23}$ that
yielded triangle inequality violations. Accordingly, if one or more
of the branch reactances is less than zero, a violation may occur.

The extension of these analytical results to the ac power flow case
remains for future work. However, in order to explore the triangle
inequality conditions for a real power system we drew a random sample
of $10^{6}$ node triplets from the Eastern Interconnect $\mathbf{E}$
matrix and tested against (\ref{eq:tri_ineq}). This empirical evaluation
showed that the triangle inequality holds for $99.81\%$ of the node
triplets. 

\section*{Acknowledgment}

This work is supported in part by the US National Science Foundation
(award \#0848247) and in part by the US Department of Energy (award
\#DE-OE0000447). Special thanks are due to Mahendra Patel, and other
staff members at PJM Applied Solutions, for providing substantial
technical advice related to this effort. 

\bibliographystyle{IEEEtran}
\bibliography{GridStructure}

% Generated by IEEEtran.bst, version: 1.13 (2008/09/30)
\begin{thebibliography}{10}
\providecommand{\url}[1]{#1}
\csname url@samestyle\endcsname
\providecommand{\newblock}{\relax}
\providecommand{\bibinfo}[2]{#2}
\providecommand{\BIBentrySTDinterwordspacing}{\spaceskip=0pt\relax}
\providecommand{\BIBentryALTinterwordstretchfactor}{4}
\providecommand{\BIBentryALTinterwordspacing}{\spaceskip=\fontdimen2\font plus
\BIBentryALTinterwordstretchfactor\fontdimen3\font minus
  \fontdimen4\font\relax}
\providecommand{\BIBforeignlanguage}[2]{{%
\expandafter\ifx\csname l@#1\endcsname\relax
\typeout{** WARNING: IEEEtran.bst: No hyphenation pattern has been}%
\typeout{** loaded for the language `#1'. Using the pattern for}%
\typeout{** the default language instead.}%
\else
\language=\csname l@#1\endcsname
\fi
#2}}
\providecommand{\BIBdecl}{\relax}
\BIBdecl

\bibitem{Erdos:1959}
P.~Erd{\"o}s and A.~R{\'e}nyi, ``On random graphs,'' \emph{Publ. Math.
  Debrecen}, vol.~6, pp. 290--297, 1959.

\bibitem{Barabasi:1999}
A.~Barab{\'a}si and R.~Albert, ``Emergence of scaling in random networks,''
  \emph{Science}, vol. 286, no. 5439, pp. 509--512, 1999.

\bibitem{Watts:1998}
D.~J. Watts and S.~H. Strogatz, ``Collective dynamics of 'small-world'
  networks,'' \emph{Nature}, vol. 393, no. 6684, pp. 440--442, 1998.

\bibitem{Boccaletti:2006}
S.~Boccaletti, V.~Latora, Y.~Moreno, M.~Chavez, and D.~Hwang, ``Complex
  networks: Structure and dynamics,'' \emph{Physics reports}, vol. 424, no.
  4-5, pp. 175--308, 2006.

\bibitem{Albert:2000}
R.~Albert, H.~Jeong, and A.~Barab{\'a}si, ``Error and attack tolerance of
  complex networks,'' \emph{Nature}, vol. 406, no. 6794, pp. 378--382, 2000.

\bibitem{Wang:2002}
X.~Wang and G.~Chen, ``Synchronization in scale-free dynamical networks:
  robustness and fragility,'' \emph{IEEE Transactions on Circuits and Systems
  I: Fundamental Theory and Applications}, vol.~49, no.~1, pp. 54--62, 2002.

\bibitem{Li:2004}
X.~Li, X.~Wang, and G.~Chen, ``Pinning a complex dynamical network to its
  equilibrium,'' \emph{IEEE Transactions on Circuits and Systems I: Regular
  Papers}, vol.~51, no.~10, pp. 2074--2087, 2004.

\bibitem{Chen:2007}
T.~Chen, X.~Liu, and W.~Lu, ``Pinning complex networks by a single
  controller,'' \emph{IEEE Transactions on Circuits and Systems I: Regular
  Papers}, vol.~54, no.~6, pp. 1317--1326, 2007.

\bibitem{Atay:2006}
F.~Atay, T.~Biyikoglu, and J.~Jost, ``Synchronization of networks with
  prescribed degree distributions,'' \emph{IEEE Transactions on Circuits and
  Systems I: Regular Papers}, vol.~53, no.~1, pp. 92--98, 2006.

\bibitem{Wu:1995}
C.~Wu and L.~Chua, ``Synchronization in an array of linearly coupled dynamical
  systems,'' \emph{IEEE Transactions on Circuits and Systems I: Fundamental
  Theory and Applications}, vol.~42, no.~8, pp. 430--447, 1995.

\bibitem{Wu:2005}
C.~Wu, ``Synchronization in networks of nonlinear dynamical systems coupled via
  a directed graph,'' \emph{Nonlinearity}, vol.~18, p. 1057, 2005.

\bibitem{De-Lellis:2008}
P.~De~Lellis, M.~di~Bernardo, and F.~Garofalo, ``Synchronization of complex
  networks through local adaptive coupling,'' \emph{Chaos: An Interdisciplinary
  Journal of Nonlinear Science}, vol.~18, no.~3, p. 037110, 2008.

\bibitem{De-Lellis:2010a}
P.~De~Lellis, M.~di~Bernardo, F.~Garofalo, and M.~Porfiri, ``Evolution of
  complex networks via edge snapping,'' \emph{IEEE Transactions on Circuits and
  Systems I: Regular Papers}, vol.~57, no.~8, pp. 2132--2143, 2010.

\bibitem{Dorfler:2010}
F.~Dorfler and F.~Bullo, ``Synchronization and transient stability in power
  networks and non-uniform kuramoto oscillators,'' in \emph{American Control
  Conference (ACC), Baltimore, MD, USA}.\hskip 1em plus 0.5em minus 0.4em\relax
  IEEE, 2010, pp. 930--937.

\bibitem{Dobson:2007}
I.~Dobson, B.~Carreras, V.~Lynch, and D.~Newman, ``{Complex systems analysis of
  series of blackouts: Cascading failure, critical points, and
  self-organization},'' \emph{Chaos: An Interdisciplinary Journal of Nonlinear
  Science}, vol.~17, no.~2, p. 026103, 2007.

\bibitem{Amaral:2000}
L.~Amaral, A.~Scala, M.~Barth{\'e}l{\'e}my, and H.~Stanley, ``Classes of
  small-world networks,'' \emph{Proceedings of the National Academy of Sciences
  of the United States of America}, vol.~97, no.~21, pp. 11\,149--11\,152,
  2000.

\bibitem{Albert:2004}
R.~Albert, I.~Albert, and G.~Nakarado, ``Structural vulnerability of the north
  american power grid,'' \emph{Physical Review E}, vol.~69, no.~2, 2004.

\bibitem{Crucitti:2004}
P.~Crucitti, V.~Latora, and M.~Marchiori, ``A topological analysis of the
  italian electric power grid,'' \emph{Physica A: Statistical Mechanics and its
  Applications}, vol. 338, no. 1-2, pp. 92--97, 2004.

\bibitem{Chassin:2005}
D.~Chassin and C.~Posse, ``Evaluating north american electric grid reliaiblity
  using the barabasi-albert network model,'' \emph{Physica A: Statistical
  Mechanics and its Applications}, vol. 355, no. 2-4, pp. 667--677, 2005.

\bibitem{Holmgren:2006}
{\AA}.~Holmgren, ``Using graph models to analyze the vulnerability of electric
  power networks,'' \emph{Risk Analysis}, vol.~26, no.~4, pp. 955--969, 2006.

\bibitem{Sole:2008}
R.~Sol{\'e}, M.~Rosas-Casals, B.~Corominas-Murtra, and S.~Valverde,
  ``Robustness of the european power grids under intentional attack,''
  \emph{Physical Review E}, vol.~77, no.~2, p. 026102, 2008.

\bibitem{Wang:2010}
Z.~Wang, A.~Scaglione, and R.~Thomas, ``Generating statistically correct random
  topologies for testing smart grid communication and control networks,''
  \emph{IEEE Transactions on Smart Grid}, vol.~1, no.~1, pp. 28--39, 2010.

\bibitem{Wang:2009}
J.~Wang and L.~Rong, ``Cascade-based attack vulnerability on the us power
  grid,'' \emph{Safety Science}, 2009.

\bibitem{Hines:2010b}
P.~Hines, E.~Cotilla-Sanchez, and S.~Blumsack, ``Do topological models provide
  good information about vulnerability in electric power networks?''
  \emph{Chaos: An Interdisciplinary Journal of Nonlinear Science}, vol.~20,
  no.~3, p. 033122, 2010.

\bibitem{Wang:2010a}
Z.~Wang, A.~Scaglione, and R.~Thomas, ``Electrical centrality measures for
  electric power grid vulnerability analysis,'' in \emph{Proceedings of the
  49th IEEE Conference on Decision and Control, Atlanta, Georgia}.\hskip 1em
  plus 0.5em minus 0.4em\relax IEEE, 2010, pp. 5792--5797.

\bibitem{Hines:2008}
P.~Hines and S.~Blumsack, ``A centrality measure for electrical networks,'' in
  \emph{Proceedings of the 41st Hawaii International Conference on System
  Sciences}, 2008.

\bibitem{Hines:2010a}
P.~Hines, S.~Blumsack, E.~Cotilla-Sanchez, and C.~Barrows, ``The topological
  and electrical structure of power grids,'' in \emph{Proceedings of the 43rd
  Hawaii International Conference on System Sciences}, 2010.

\bibitem{Bompard:2009}
E.~Bompard, R.~Napoli, and F.~Xue, ``Analysis of structural vulnerabilities in
  power transmission grids,'' \emph{International Journal of Critical
  Infrastructure Protection}, vol.~2, no. 1-2, pp. 5--12, 2009.

\bibitem{Arianos:2009}
S.~Arianos, E.~Bompard, A.~Carbone, and F.~Xue, ``Power grids vulnerability: a
  complex network approach,'' \emph{Chaos: An Interdisciplinary Journal of
  Nonlinear Science}, vol.~19, no.~1, p. 013119, 2009.

\bibitem{Blumsack:2007a}
S.~Blumsack, L.~Lave, and M.~Ilic, ``A quantitative analysis of the
  relationship between congestion and reliability in electric power networks,''
  \emph{Energy Journal}, vol.~28, pp. 101--128, 2007.

\bibitem{Ming:2006}
D.~Ming and H.~Ping-ping, ``Small-world topological model based vulnerability
  assessment algorithm for large-scale power grid,'' \emph{Automation of
  Electric Power Systems}, vol.~8, 2006.

\bibitem{Rosato:2007}
V.~Rosato, S.~Bologna, and F.~Tiriticco, ``Topological properties of
  high-voltage electrical transmission networks,'' \emph{Electric power systems
  research}, vol.~77, no.~2, pp. 99--105, 2007.

\bibitem{Fu:2010}
L.~Fu, W.~Zhang, S.~Xiao, Y.~Li, and S.~Guo, ``Vulnerability assessment for
  power grid based on small-world topological model,'' in \emph{Power and
  Energy Engineering Conference (APPEEC), 2010 Asia-Pacific}, 2010.

\bibitem{Rosas-Casals:2007}
M.~Rosas-Casals, S.~Valverde, and R.~Sol{\'e}, ``Topological vulnerability of
  the european power grid under errors and attacks,'' \emph{International
  Journal of Bifurcation and Chaos}, vol.~17, no.~7, pp. 2465--2475, 2007.

\bibitem{PSTCA:2007}
\BIBentryALTinterwordspacing
PSTCA. (2007) {Power Systems Test Case Archive, University of Washington,
  Electrical Engineering}. [Online]. Available:
  \url{http://www.ee.washington.edu/research/pstca/}
\BIBentrySTDinterwordspacing

\bibitem{Albert:2002}
R.~Albert and A.~Barab{\'a}si, ``Statistical mechanics of complex networks,''
  \emph{Reviews of modern physics}, vol.~74, no.~1, p.~47, 2002.

\bibitem{Newman:2002}
M.~Newman, ``Assortative mixing in networks,'' \emph{Physical Review Letters},
  vol.~89, no.~20, p. 208701, 2002.

\bibitem{Newman:2003a}
------, ``Mixing patterns in networks,'' \emph{Physical Review E}, vol.~67,
  no.~2, p. 026126, 2003.

\bibitem{Clauset:2009}
A.~Clauset, C.~Shalizi, and M.~Newman, ``Power-law distributions in empirical
  data,'' \emph{SIAM Review}, vol.~51, no.~4, pp. 661--703, 2009.

\bibitem{Zimmerman:2011}
R.~D. Zimmerman, C.~E. Murillo-S{\'a}nchez, and R.~J. Thomas, ``{MATPOWER}:
  Steady-state operations, planning and analysis tools for power systems
  research and education,'' \emph{IEEE Transactions on Power Systems}, vol.~26,
  no.~1, pp. 12--19, 2011.

\bibitem{Bialek:1997}
J.~Bialek, ``Topological generation and load distribution factors for
  supplement charge allocation in transmission open access,'' \emph{IEEE
  Transactions on Power Systems}, vol.~12, no.~3, pp. 1185--1193, 1997.

\bibitem{Lagonotte:1989}
P.~Lagonotte, J.~Sabonnadiere, J.~Leost, and J.~Paul, ``Structural analysis of
  the electrical system: application to secondary voltage control in france,''
  \emph{IEEE Transactions on Power Systems}, vol.~4, no.~2, pp. 479--486, 1989.

\bibitem{Lu:1995}
Q.~Lu and S.~Brammer, ``A new formulation of generator penalty factors,''
  \emph{IEEE Transactions on Power Systems}, vol.~10, no.~2, pp. 990--994,
  1995.

\bibitem{Hang:2000}
L.~Hang, A.~Bose, and V.~Venkatasubramanian, ``A fast voltage security
  assessment method using adaptive bounding,'' \emph{IEEE Transactions on Power
  Systems}, vol.~15, no.~3, pp. 1137--1141, 2000.

\bibitem{Zhong:2004}
J.~Zhong, E.~Nobile, A.~Bose, and K.~Bhattacharya, ``Localized reactive power
  markets using the concept of voltage control areas,'' \emph{IEEE Transactions
  on Power Systems}, vol.~19, no.~3, pp. 1555--1561, 2004.

\bibitem{Klein:1993}
D.~Klein and M.~Randic, ``Resistance distance,'' \emph{Journal of Mathematical
  Chemistry}, vol.~12, pp. 81--95, 1993.

\bibitem{Dobson:2010b}
I.~Dobson, M.~Parashar, and C.~Carter, ``Combining phasor measurements to
  monitor cutset angles,'' in \emph{Proc. of the 43rd Hawaii International
  Conference on System Sciences}, Kauai, HI, 2010.

\bibitem{Rudin:1964}
W.~Rudin, \emph{Principles of mathematical analysis}.\hskip 1em plus 0.5em
  minus 0.4em\relax McGraw-Hill New York, 1964, vol. 1976.

\bibitem{Stott:2009}
B.~Stott, J.~Jardim, and O.~Alsac, ``{DC} power flow revisited,'' \emph{IEEE
  Transactions on Power Systems}, vol.~24, no.~3, pp. 1290--1300, 2009.

\end{thebibliography}

\end{document}